\documentclass[aps,prd,twocolumn,groupedaddress,nofootinbib]{revtex4-1}
\usepackage{amsfonts,ulem,bbold}
\usepackage{amssymb,amsmath}
\usepackage{color}
\usepackage{epsfig}

\usepackage{breakurl}

\newcommand{\Comment}[1]{{}}
\definecolor{MyDarkBlue}{rgb}{0.15,0.15,0.45}
\usepackage[linktocpage=true]{hyperref}
\hypersetup{
colorlinks=true,
citecolor=MyDarkBlue,
linkcolor=MyDarkBlue,
urlcolor=MyDarkBlue,
}

\newcommand{\be}{\begin{equation}}
\newcommand{\ee}{\end{equation}}
\newcommand{\bea}{\begin{eqnarray}}
\newcommand{\eea}{\end{eqnarray}}

\begin{document}

\title{Gravitational Collapse in Massive Gravity on de Sitter Spacetime}

\author{Roman Berens}
\affiliation{Department of Physics, Columbia University,\\ New York, NY 10027, USA}
\author{Luke Krauth}
\affiliation{Department of Physics, Columbia University,\\ New York, NY 10027, USA}
\author{Rachel A. Rosen}
\affiliation{Department of Physics, Columbia University,\\ New York, NY 10027, USA}

\begin{abstract} 

We analyze the evolution of a homogenous and pressureless ball of dust (or ``star") in ghost-free massive gravity on de Sitter spacetime.  We find that gravitational collapse does not take place for all parameters of the massive gravity theory.  For parameters where it does occur, we find the expression for the location of the apparent horizon where it crosses the surface of the star, indicating the location of the apparent horizon of the vacuum solution at that moment.  We determine the Ricci curvature at the boundary of the star and extract the finite correction to the curvature of the apparent horizon due to the graviton mass.  Finally, we argue that our collapsing solutions cannot be matched to a static, spherically symmetric vacuum solution at the star's surface, providing further evidence that physical black hole solutions in massive gravity are likely time-dependent.

\end{abstract}

\maketitle

\section{Introduction and Summary}
\label{intro}
 \vspace{-.5cm}
Black holes solutions in massive gravity are interesting for a variety of reasons.  They can serve as an alternative benchmark model as we test General Relativity to higher and higher precision using black hole physics.   Moreover, understanding black hole thermodynamics when the graviton has a mass might offer a new perspective on black hole information.  Yet the determination of physical, analytic black hole solutions in ghost-free massive gravity remains a difficult problem.  A central issue is that static black hole solutions are either infinitely strongly coupled and unphysical, or inevitably possess a coordinate-invariant curvature singularity at the horizon \cite{Deffayet:2011rh,Mirbabayi:2013sva,Rosen:2017dvn}.  Perturbative, time-dependent solutions can be found that alleviate the curvature singularity \cite{Rosen:2017dvn}.  However, massive gravity allows for many branches of solutions and it remains unclear what the correct physical branch is.  Another open question is whether black holes can even be formed within the regime of validity of the massive gravity effective field theory, which possesses a notably low cutoff.

In this paper we attempt to shed light on these issues by studying Oppenheimer-Snyder collapse \cite{Oppenheimer:1939ue} of a homogenous, pressureless ball of dust (or ``star") in ghost-free massive gravity on de Sitter spacetime.  We consider a range of parameter space and ask whether collapse takes place and black hole formation is possible.  For the collapsing solutions, we locate the position of the apparent horizon where it crosses the surface of the star, indicating the location of the apparent horizon for the vacuum solution at that moment.  We determine the finite contribution to the Ricci curvature at the apparent horizon due to the graviton mass.  We find that it is the ``minimal model" of massive gravity in de Sitter spacetime that exhibits collapsing solutions that recover those of General Relativity in the limits of small graviton mass and flat spacetime.  We then show that these collapsing solutions cannot be matched to static, spherically symmetric vacuum solutions on the exterior.  This provides support to the argument that physical black holes in massive gravity are likely time-dependent.

\section{Massive Gravity in de Sitter}
 \vspace{-.5cm}
Our starting point is dRGT ghost-free massive gravity \cite{deRham:2010kj} with a de Sitter reference metric $f_{\mu\nu}$ with Hubble constant $H$ and a dynamical metric $g_{\mu\nu}$ with a cosmological constant $\Lambda = 3 H^2$:  
\be
\label{L}
{\cal L} = \frac{M_{P}^2}{2}\sqrt{-g}\left[R[g]-6H^2-2 m^2\sum_{n=0}^4 \beta_n S_n(\sqrt{g^{-1} f}) \right] \, .
\ee
In the potential term, the $S_n$ are the $n$-th elementary symmetric polynomials of the eigenvalues of the matrix square root of $g^{\mu\lambda}f_{\lambda\nu}$.  They are given by
\bea
\label{potential}
\begin{array}{l}
S_0 (\mathbb{X})= 1  \, ,  \vspace{.1cm} \\
S_1(\mathbb{X})= [\mathbb{X}]  \, ,  \vspace{.1cm} \\
S_2(\mathbb{X})= \tfrac{1}{2}([\mathbb{X}]^2-[\mathbb{X}^2]) \, ,  \vspace{.1cm} \\
S_3(\mathbb{X})= \tfrac{1}{6}([\mathbb{X}]^3-3[\mathbb{X}][\mathbb{X}^2]+2[\mathbb{X}^3]) \, ,  \vspace{.1cm} \\
S_4(\mathbb{X})=\tfrac{1}{24}([\mathbb{X}]^4-6[\mathbb{X}]^2[\mathbb{X}^2]+3[\mathbb{X}^2]^2  +8[\mathbb{X}][\mathbb{X}^3]-6[\mathbb{X}^4])\, ,
\end{array}
\eea
where here (and only here) the square brackets denote the trace of the enclosed matrix.   This form of the potential guarantees that the classical theory propagates only the correct five degrees of freedom of the massive graviton \cite{Hassan:2011hr,Hassan:2011tf} and no additional Boulware-Deser ghost \cite{Boulware:1973my}.

The $\beta_n$ are free constant coefficients subject to two constraints.  The requirement of no tadpoles when expanding $g_{\mu\nu}$ about $f_{\mu\nu}$ fixes
\be
\label{cc}
\beta_0 + 3 \beta_1 + 3 \beta_2 + \beta_3 = 0 \, ,
\ee
while the correct normalization of the Fierz-Pauli mass $m^2$  fixes
\be
\label{mass}
\beta_1 +2\beta_2 +\beta_3 =1 \, .
\ee
In addition, $\beta_4$ multiplies a term which is non-dynamical and doesn't contribute to the equations of motion.  Thus, of the five $\beta_n$ there are two independent parameters, in addition to the Planck mass $M_P$, the graviton mass $m$ and the Hubble constant $H$.  In what follows we will use $\beta_2$ and $\beta_3$ to parametrize these remaining two free parameters.

To aid our analysis, we divide parameter space into the following regions and identify special points within the regions:

\begin{itemize}

\item{{\it The Minimal Model:  $\beta_2 = \beta_3 = 0$.}  This choice of parameters correspond to $c_3 = \tfrac{1}{6}$ and $d_5 = -\tfrac{1}{48}$ in the parametrization of \cite{deRham:2010ik}.  In flat spacetime all interactions of the helicity-0 mode vanish in the decoupling limit at this point.  Static black holes with flat asymptotics have no Vainshtein mechanism \cite{Vainshtein:1972sx} in the minimal model, as was shown in \cite{Renaux-Petel:2014pja,Rosen:2017dvn}, and thus do not recover General Relativity in the limit of small graviton mass.  In contrast, in de Sitter spacetime we will see that the minimal model does have a smooth limit with General Relativity.}

\item{{\it The Next-to-Minimal Model: $\beta_2 \neq 0$ and $\beta_3 = 0$.}   In de Sitter spacetime and at the Higuchi bound  $m^2 = 2 H^2$ \cite{Higuchi:1986py}, this set of parameters includes the candidate non-linear ``partially massless"  \cite{Deser:1983mm,Deser:2001pe,Deser:2001us}  model with $\beta_2 = \tfrac{1}{2}$ and $\beta_3 = 0$.  This theory was identified in \cite{deRham:2012kf} and studied in \cite{deRham:2013wv}.  In \cite{DeRham:2018axr}, this model corresponds to the parameters $\alpha_3 = -\tfrac{1}{2}$ and $\alpha_4 = \tfrac{1}{8}$ and it was shown to have a high strong coupling scale compared to the generic value.}

\item{{\it The Non-Minimal Model: $\beta_2 \neq 0$ and $\beta_3 \neq 0$.}  This set of parameters includes the ``${\mathbb Z}_2$ model" with $\beta_0 = \beta_4$ and $\beta_1 = \beta_3$.  %For this choice, the $\beta_n$ exhibit a ${\mathbb Z}_2$ symmetry under the interchange of $\beta_n \leftrightarrow \beta_{4-n}$.  
In flat spacetime, the ${\mathbb Z}_2$ model coincides with the model found to avoid asymptotic superluminality \cite{Hinterbichler:2017qyt}, i.e., it corresponds to $c_3 = \tfrac{1}{4}$ in the parametrization used there.  In the parametrization of \cite{DeRham:2018axr}, this model corresponds to $\alpha_3 = -\tfrac{1}{2}$.  There, it was found that at this point in parameter space, in de Sitter spacetime and at  $m^2 = 2 H^2$, the strong coupling scale of the theory is also raised above its generic value.  }

\end{itemize}

In what follows, we will consider the cases $\beta_3 = 0$ and $\beta_3 \neq 0$ separately, as well as the special points they include.

\section{Collapsing Star}
 \vspace{-.5cm}
Let us consider a pressureless and homogenous ball of dust.  We take the metric of the interior to be homogenous and isotropic and we use conformal time $\eta$ and comoving radial coordinate $r$, so that the dynamical metric is given by
\be
\label{gmetric}
g_{\mu\nu} dx^\mu dx^\nu = a(\eta)^2 \left(-d\eta^2 +\frac{dr^2}{1-k^2 r^2}+r^2d\Omega^2 \right) \, ,
\ee
where $d\Omega^2 = d\theta^2+\sin^2\theta\, d\phi^2$.  The function $a(\eta)$ will be determined by the equations of motion and the spatial curvature scale $k$ will be determined by the density of the dust, the de Sitter curvature scale $H$, and the graviton mass $m$.

We express the de Sitter reference metric in terms of a different conformal time $\tau$ and comoving radius $\rho$ :
\be
\label{fmetric}
f_{\mu\nu} dx^\mu dx^\nu = \frac{1}{\cos(H \tau)^2} \left(-d\tau^2 +\frac{d\rho^2}{1-H^2 \rho^2}+\rho^2d\Omega^2 \right) \, .
\ee
In general, $(\tau,\rho)$ need not coincide with $(\eta,r)$ and we can have $\tau=\tau(\eta,r)$ and $\rho = \rho(\eta,r)$.  These two functions will also be fixed by the equations of motion.

Our source is a pressureless, spherically symmetric ball of dust of uniform density $\mu(\eta)$:
\be
\label{T}
T^{\mu\nu} = \mu(\eta)\, u^\mu u^\nu \, ,
\ee
with
\be
u^\mu = \frac{1}{a(\eta)} \left\{1,0,0,0\right\} \, .
\ee
We set our clocks so that the scale factor $a = 1$ when $\eta = 0$.  We can thus express the energy density as
\be
\mu(\eta) = \frac{\mu_0}{a(\eta)^3}\, ,
\ee
with $\mu_0$ being the initial density.  With these expressions, the conservation of the stress tensor is automatically satisfied: $\nabla^\mu T_{\mu\nu}=0$.  

For later convenience we define the quantity $M$ so that
\be
\label{M}
M = \tfrac{4}{3} \pi r_0^3 *\mu_0 \, ,
\ee
where $r_0$ is the comoving radius of the ball of dust which is equal to the physical radius when $\eta = 0$.

\section{Equations of Motion}
 \vspace{-.5cm}
We assume the massive graviton is minimally coupled to the source \eqref{T}.  The equations of motion that follow from the Lagrangian \eqref{L} can be written as
\be
\label{EOM}
R_{\mu\nu} -\tfrac{1}{2}R\,g_{\mu\nu}+3H^2g_{\mu\nu}+m^2 U_{\mu\nu} = 8 \pi G\, T_{\mu\nu}\, ,
\ee
where $U_{\mu\nu}$ is the term that follows from the massive graviton potential.  We use the above expressions for $g_{\mu\nu}$ \eqref{gmetric} and $f_{\mu\nu}$ \eqref{fmetric} in the equations of motion.  In particular, there are three unknown functions $a(\eta)$, $\tau(\eta,r)$ and $\rho(\eta,r)$ that need to be satisfied by the three independent equations that arise from \eqref{EOM}.  We note that our setup means that we are effectively looking for cosmological solutions in massive gravity on de Sitter and our analysis of equation \eqref{EOM} has overlap with the analysis of \cite{Langlois:2012hk}.

We find solutions of the following form:
\be
\label{tr}
\tau(\eta,r)=\tau(\eta) \, , ~~~~\rho(\eta,r) = \frac{k}{H} r \, .  
\ee
That such solutions exist for a homogenous and isotropic ansatz \eqref{gmetric} is directly due to the fact that we are in de Sitter spacetime and thus both $g$- and $f$-metrics have slicings with positive spatial curvature.  

We are left with two independent equations for $a(\eta)$ and $\tau(\eta)$.  By applying the Bianchi identity to equation \eqref{EOM} we find two possible branches of solutions for $\tau(\eta)$:\\
\\
{\it BRANCH I:}
\be
\label{branch1}
H^2 \beta_1 a(\eta)^2 +  2 H k \beta_2 a(\eta) \sec\left[H \tau (\eta)\right] + k^2 \beta_3 \sec\left[H \tau(\eta)\right]^2
=0  \, ,
\ee
{\it BRANCH II:}
\be
\label{branch2}
\tan\left[H \tau(\eta)\right] = \frac{a'(\eta)}{k\, a(\eta)} \, ,
\ee
where primes denote derivatives with respect to conformal time $\eta$.

The first branch is the usual pathological branch where exact solutions can be found (see, e.g., \cite{Mirbabayi:2013sva,Rosen:2017dvn}).  With the source turned off, i.e., $\mu_0 = 0$, the $f$- and $g$-metrics are the same, and we have $k=H$, $\tau = \eta$ and $a(\eta)  = \sec (H \eta)$.  Thus, equation \eqref{branch1} means that this branch requires
\be
\beta_1 +2\beta_2 +\beta_3 = 0 \, ,
\ee
away from sources.  This is in contradiction to the constraint \eqref{mass} and effectively sets the Fierz-Pauli mass to zero.  On this branch one finds identical solutions to those of General Relativity but, because the mass vanishes around a de Sitter background, they are infinitely strongly coupled and unphysical.

We thus focus our attention on the second branch of solutions \eqref{branch2}.  We are left with one remaining equation of motion for $a(\eta)$:
\begin{multline}
\label{aeom}
3\left(\frac{a'^2}{a^4}+\frac{k^2}{a^2}\right) =\frac{8\pi G\,\mu_0}{a^3} +3H^2\\
+m^2\left( \beta_0+3 \beta_1 {\cal F}[a,a']+ 3 \beta_2 {\cal F}[a,a']^2 +\beta_3 {\cal F}[a,a']^3  \right) \, ,
\end{multline}
where
\be
{\cal F}[a,a'] = \frac{k }{H a} \sqrt{1+\frac{a'^2}{k^2 a^2}} \, .
\ee
We can find $k$ in terms of $\mu_0$ and $H$ by setting our initial conditions.  In particular, we take 
\be
\label{init}
a(0)=1 ~~~ {\rm and} ~~~ a'(0)=0  \, ,
\ee
so that our clock begins at the moment of collapse.  The constants $\mu_0$ and $k$ are thus related by
\begin{multline}
\label{k}
k^2 =\frac{8\pi G}{3} \mu_0+H^2\\
+\frac{m^2}{3}\left( \beta_0+3 \beta_1\frac{k}{H}+ 3 \beta_2 \frac{k^2}{H^2} +\beta_3\frac{k^3}{H^3}   \right) \, .
\end{multline}

Let us consider some implications of the expressions \eqref{aeom} and \eqref{k}.   We see that taking $m  \rightarrow 0$ while holding $H$ and all other parameters fixed smoothly recovers the General Relativistic expressions for $a(\eta)$ and $k$, for generic values of the $\beta_n$ coefficients.  Though naively obvious, this is actually a non-trivial consequence of working in de Sitter spacetime.  In flat spacetime, for the minimal model ($\beta_2=\beta_3=0$), it is known that the Vainshtein mechanism doesn't hold and that the massless limit is discontinuous with General Relativity \cite{Renaux-Petel:2014pja,Rosen:2017dvn}.

Alternatively, we can consider the flat space limit at finite graviton mass.  We see that terms arising from the massive graviton potential generically blow up when taking $H \rightarrow 0$ while keeping $m$ and other parameters fixed.  It is reasonable, however, to consider a situation in which the scale of the graviton mass and the de Sitter scale are the same: $m \sim H$.  This is often the set-up in cosmological scenarios and is consistent with observational bounds on the gravitational mass (see, e.g., \cite{deRham:2016nuf}).  It also arises naturally from the theory itself: a cosmological constant can always be absorbed into the normalization of the $\beta_n$ coefficients and is thus set by the same scale as the graviton mass.  If we now consider taking the combined limit  $m,H \rightarrow 0$, we see that only the $\beta_3$ term blows up in the above expressions.  

For physical collapse, it is reasonable to assume that $k \gg H,m$, i.e., that the spatial curvature radius of the collapsing ball of dust is much shorter than the de Sitter radius or the wavelength of the massive graviton.  We thus find the constraint $\beta_3 \leq 0 $ is required for consistency with $\mu_0$ being positive.  Setting $\beta_3 = 0$ and taking $m,H \rightarrow 0$ we see the $\beta_2$ term adds a finite, order 1 correction to the General Relativistic expressions.  The positivity of $\mu_0$ now enforces $\beta_2 \leq \tfrac{H^2}{m^2}$.  At the Higuchi bound $m^2 = 2 H^2$, this gives $\beta_2 \leq \tfrac{1}{2}$.    Finally, with $\beta_2=0$ and $\beta_3=0$, the final non-trivial term $\beta_1$ gives a perturbative correction to the general relativistic expressions that vanishes as $m,H \rightarrow 0$.  We see that the minimal model in de Sitter spacetime will recover General Relativity in flat spacetime when $m,H \rightarrow 0$.

To solve \eqref{aeom} for $a(\eta)$, we first determine $a'$ as a function of $a$.  We express this as solution as
\be
\label{f}
a'=-k\sqrt{f[a]} \, ,
\ee
where we have chosen the root with the overall negative sign as we are interested in the collapsing solution.  In general, for generic $\beta_n$, there are six roots of the expression \eqref{aeom} and thus multiple possible $f[a]$.  Only one root will correspond to a physical solution consistent with the appropriate initial conditions \eqref{init} and the parameters of the theory.  Then, $a(\eta)$ is determined by inverting the expression
\be
\eta(a) = -\frac{1}{k}\, \int_1^a \frac{d{\tilde a}}{\sqrt{f[\tilde{a}]}} \, . 
\ee
Using the above expression, the collapse time is given by $\eta_{collapse} = \eta(0)$.
%For zero graviton mass and in flat spacetime, one recovers the usual expression
%\be
%a(\eta) = \tfrac{1}{2}(1+\cos[k \eta]) \,.
%\ee
%I.e., the star collapses from finite comoving radius $r_0$ to zero in finite conformal time $\eta_{collapse} = \pi/k$.  

In de Sitter spacetime and at nonzero graviton mass, the expressions for $a(\eta)$ and $\eta_{collapse}$ are complicated.  In general, only perturbative results can be obtained.  However, the expression for $f[a]$ as defined in \eqref{f} is all we need to extract other physical results exactly, including the polynomial expression for the location of the apparent horizon and the expression for the Ricci curvature inside the star.  By considering matching at the star's boundary, we can extract non-perturbative statements about the vacuum solutions, i.e., about black holes in massive gravity.  We perform this analysis in the next section.

\section{Apparent Horizons}
\label{horizons}
 \vspace{-.5cm}
An important question is how the mass of the graviton affects the location of the black hole horizon and whether it is possible that graviton mass prevents the formation of a horizon entirely.  To consider this, we first perform a coordinate transformation $(\eta,r) \rightarrow (T,R)$ to put the dynamical metric \eqref{gmetric} in the form
\be
\label{gin}
g_{\mu\nu} dx^\mu dx^\nu = -V_0(T,R)dT^2 +\frac{dR^2}{V_1(T,R)}+R^2 d\Omega^2  \, .
\ee
Here, we have defined our new coordinates $(T,R)$ so that the new metric is diagonal and so that the area of a sphere gives the expected result when expressed in terms of $R$.  This immediately relates $R$ and $r$:
\be
R = a(\eta)\, r \, .
\ee
Furthermore, starting from \eqref{gmetric}, one finds that the function $V_1(T,R)$ is given by
\be
\label{V1}
V_1(T,R) = 1-\frac{k^2 R^2}{a^2}\left(1+\frac{f[a]}{a^2}  \right) \, ,
\ee
where $a = a[\eta(T,R)]$ and $f[a]$ is defined by \eqref{f}.

The location where $V_1(T,R)$ vanishes gives the location of the apparent horizon inside the ball of dust.  I.e., we can use the expression
\be
V_1(T,R_{hor}(T)) = 0\, , 
\ee
to define $R_{hor}(T)$. (See, e.g., \cite{Nielsen:2005af} for a discussion of apparent horizons in spherically symmetric, time-dependent spacetimes.) Of interest is the location of the apparent horizon where it coincides with the surface of the collapsing ball of dust, given by $R_0 = a(\eta) r_0$ in our new coordinates.  Because the interior metric must match the vacuum solution outside the star, this gives the location of the apparent horizon for the vacuum solution, at least at the moment where it coincides with the surface of the star.  In General Relativity, where the vacuum solution is static, the location where the apparent horizon crosses the surface of the star in Oppenheimer-Snyder collapse coincides with the event horizon.

To find the location of the apparent horizon at surface crossing we first find $V_1$ at the surface of the star, i.e., at $R_0= a(\eta) r_0$.  We then set $V_1 = 0$ and solve for $R_{hor}$.  Below we work out the solution for generic regions in parameter space.

\vspace{.5cm}
\noindent {\bf General Relativity:}  As both a warmup exercise and a reference, we first consider the case that $m=0$ and $H \neq 0$.  From \eqref{aeom} and \eqref{f}, we find
\be
\label{GRf}
f[a]= a\left(1-\frac{H^2}{k^2}-a+\frac{H^2}{k^2}a^3 \right) \, ,
\ee
and, from \eqref{k},
\be
\label{GRk}
k=\sqrt{\frac{2\,GM}{r_0^3}+H^2} \, ,
\ee
where we have replaced $\mu_0$ with $M$ using \eqref{M}.  Evaluating at $R_0 = a(\eta) r_0$, expression \eqref{V1} becomes
\be
V_1(R_0) = 1-\frac{2\,GM}{R_0} - H^2 R_0^2 \, ,
\ee
as we would expect for matching to an exterior Schwarzschild-de Sitter spacetime.  For later comparison, we note that solving perturbatively to find the apparent horizon at surface crossing $V_1(R_0\! = \!R_{hor}) = 0$ gives
\be
\label{GRhor}
R_{hor} = 2GM \Big(1 + 4 H^2 G^2 M^2 +{\cal O} (H^4 G^4 M^4) \Big) \, .
\ee

We now consider the Ricci curvature at the star's surface
\be
{\cal R} = 6 \left(\frac{a''}{a^3}+\frac{k}{a^2} \right) \, .
\ee
Here, we must specify explicitly whether we are referring to the interior side of the star's surface $R_0^-$ or the exterior side $R_0^+$.  Unlike the metric, the curvature will be discontinuous at the star's surface because of discontinuity of the source \eqref{T}.  Using \eqref{GRf}, we can derive the Ricci curvature at the interior of the star's surface without having to solve for $a$ explicitly,
\be
{\cal R}(R_0^-) = 12 H^2 +\frac{6\,GM}{R_0^3} \, .
\ee
The first term is the usual background de Sitter curvature while the second is from the source.  Using \eqref{EOM} and the continuity of the metric at the star's surface, it is straightforward to show that curvature on the vacuum side is given by
\be
{\cal R}(R_0^+) = 12 H^2 \, ,
\ee
as expected.  In what follows we will subtract off the source contribution $6\,GM/R_0^3$ to obtain the curvature of the vacuum at the surface of the star.

 \vspace{.5cm}
\noindent {\bf Minimal and Next-to-Minimal Models: $\beta_3=0$.} We now consider the case of massive gravity.  We consider first the parameter choice $\beta_3 =0$ while keeping $\beta_2$ general.  After deriving the general solution we then consider the implications for the minimal model and the candidate partially massless model.  For the function $f[a]$ we find
\be
\label{fgen}
f[a] = a\left(c_1-a+c_2\, a^3 + c_3\,a^{3/2} \sqrt{c_1+c_4\,a^3} \right) \,,
\ee
with
\begin{align}
&c_1 = \left(1-\frac{H}{k}\right)\left(1+\frac{H}{k} \frac{H^2-(1-\beta_2)m^2}{H^2-\beta_2 m^2}\right) \, , \\
&c_2 = \frac{1}{2}\frac{H^2}{k^2} \left( 1+\frac{(H^2-(1-\beta_2)m^2)^2}{(H^2-\beta_2 m^2)^2} \right) \, , \\
&c_3 = m^2\frac{H}{k} \frac{1-2\beta_2}{H^2-\beta_2 m^2}\, , \\
&c_4 =\frac{1}{4} \frac{H^2}{k^2}\frac{(m^2-2H^2)^2}{(H^2-\beta_2 m^2)^2} \, ,
\end{align}
and
\begin{align}
\label{kgen}
k=&\frac{H}{\sqrt{H^2-\beta_2 m^2}} \sqrt{\frac{2GM}{r_0^3}+\frac{1}{4}\frac{(m^2-2H^2)^2}{H^2-\beta_2 m^2}} \nonumber \\
&+\frac{m^2}{2}\frac{H(1-2 \beta_2)}{H^2-\beta_2 m^2} \, .
\end{align}
%Consistency with initial conditions \eqref{init} requires that we  pick the positive sign in \eqref{fgen} when
%\be
%\label{cond1}
%2 k (H^2 - \beta_2  m^2) > H m^2 (1 - 2 \beta_2) \, .
%\ee
%and the negative sign in \eqref{fgen} when
%\be
%\label{cond2}
%2 k (H^2 - \beta_2  m^2) < H m^2 (1 - 2 \beta_2) \, .
%\ee
In order for collapse to occur to zero physical radius, i.e., to $a=0$, in finite time $\eta$, $f[a]$ must remain positive as $a$ goes from 1 to 0.  It is straightforward to show that this requires 
\be
k \gg \frac{m^2}{2}\frac{H(1 - 2 \beta_2)}{ (H^2 - \beta_2  m^2)} \, .
\ee
Considering \eqref{kgen}, this condition is broadly satisfied given our assumption that the radius of curvature of the star should be much shorter than the de Sitter radius and the wavelength of the graviton: $k \gg m,H$. However, if $\beta_2$ approaches the threshold value $\beta_2 = H^2/m^2$ then this condition can be violated and standard collapse will not occur.  

From the above expressions, we can derive $V_1(R_0)$ as well as ${\cal R}(R_0^+)$.  We find
\begin{widetext}
\be
\label{V1gen}
V_1(R_0) = 1-\frac{H^2}{H^2-\beta_2 m^2}\frac{2\,GM}{R_0}
 -\frac{1}{2}\left( 1+\frac{(H^2-(1-\beta_2 )m^2)^2}{(H^2-\beta_2 m^2)^2)} \right) H^2 R_0^2
 -m^2 \frac{H^2(1-2 \beta_2)}{(H^2-\beta_2m^2)^{3/2}}R_0^2\sqrt{\frac{2GM}{R_0^3}+\frac{1}{4}\frac{(m^2-2H^2)^2}{H^2-\beta_2 m^2}} \, ,
\ee
and
\be
\label{Rgen}
{\cal R}(R_0^+) = 12 H^2 
+6m^2\left( \frac{1}{2} \frac{\beta_2}{H^2-\beta_2 m^2}\frac{2GM}{R_0^3} 
+\frac{H^2(1-2 \beta_2)(m^2-2H^2)}{(H^2-\beta_2 m^2)^2}   
+\frac{1}{2}\frac{H^2(1-2\beta_2)}{(H^2-\beta_2 m^2)^{3/2}} \frac{\frac{5GM}{R_0^3}+\frac{(m^2-2H^2)^2}{H^2-\beta_2 m^2}}{\sqrt{\frac{2GM}{R_0^3}+\frac{1}{4}\frac{(m^2-2H^2)^2}{H^2-\beta_2 m^2}}}
  \right)\, .
\ee

\end{widetext}
The second term in \eqref{Rgen} is the contribution of the graviton mass term $m^2\, U_{\mu\nu}$ in \eqref{EOM} to the curvature of the vacuum at the boundary of the star.  When the apparent horizon meets the surface of the star $R_0 = R_{hor}$, this term gives the curvature at the black hole horizon arising from the graviton mass.  What's notable is that this term is finite, in contrast to the infinite curvature found at the black hole horizons for static solutions in massive gravity \cite{Deffayet:2011rh,Mirbabayi:2013sva,Rosen:2017dvn}.

It is straightforward to check that these solutions recover those of General Relativity when $m \rightarrow 0$, keeping $H$ and all other parameters fixed.  The flat space limit is less trivial.  As mentioned above, sending $H\rightarrow 0$ without also taking $m \rightarrow 0$ is a singular limit.  The limit is well-defined if we set $m^2 = 2 \lambda H^2$ for some parameter $\lambda \geq 1$ when taking $H \rightarrow 0$.  However, only in the case of the minimal model with $\beta_2=0$ do these solutions recover those of General Relativity in this limit.  Otherwise, one is left with finite corrections to the expected expressions:
\be
V_1(R_0) \rightarrow 1-\frac{2GM}{R_0} \frac{1}{1-2\lambda\beta_2 } \, ,
\ee
and
\be
{\cal R}(R_0^+) \rightarrow \frac{6GM}{R_0^3} \frac{2\lambda\beta_2 }{1-2\lambda\beta_2 } \, .
\ee
For the minimal model, we can calculate the perturbative correction to the location of the horizon due to the graviton mass.  Setting  $V_1(R_0\! = \!R_{hor}) = 0$ in \eqref{V1gen} and solving for $R_{hor}$, we find
\be
R_{hor} = 2GM \left(1 + 2 \frac{m^2}{H} G M  +{\cal O} (H^2 G^2 M^2) \right) \, ,
\ee
in contrast to \eqref{GRhor}.

%The above expressions are lengthy, so let us specialize to specific points in parameter space.  We first consider the minimal model at the Higuchi bound: $\beta_2=0$ and $m^2 = 2 H^2$.  The previous expressions simplify greatly and we find:
%\be
%\label{fmin}
%f[a] = a\left(\left(1-\frac{H}{k}\right)^2-a+\frac{H^2}{k^2}a^3 +2\frac{H}{k}\left(1-\frac{H}{k}\right)a^{3/2}\right) ,
%\ee
%and
%\be
%k=\sqrt{\frac{2\,GM}{r_0^3}}+H \, .
%\ee
%In order for collapse to occur to zero physical radius, i.e., to $a=0$, in finite time $\eta$, $f[a]$ must remain positive as $a$ goes from 1 to 0.  We see here this is always satisfied as long as $k \gg H$ as expected.

%The metric component $V_1$ at the surface of the star is given by
%\be
%V_1(R_0) = 1-\frac{2\,GM}{R_0} - H^2 R_0^2 -2H\sqrt{2GMR_0} \, .
%\ee
%Solving perturbatively for the location of the apparent horizon at surface crossing $V_1(R_0 = R_{hor}) = 0$ gives
%\be
%R_{hor} = 2GM \Big(1 + 4 H G M  +{\cal O} (H^2 G^2 M^2) \Big) \, ,
%\ee
%in contrast to \eqref{GRhor}.  We see that the effect of the graviton mass is to increase the size of the horizon.  

%The Ricci curvature at the star's surface (on the vacuum side) is now
%\be
%{\cal R}(R_0^+) = 12 H^2 +15 H \frac{\sqrt{2 G M}}{R_0^{3/2}}\, .
%\ee
%Again, the second term represents the contribution of the curvature coming from the graviton mass.

Finally, we note that the candidate partially massless model with $m^2 = 2 H^2$, $\beta_2 = \tfrac{1}{2}$ and $\beta_3 = 0$ represents a pathological point for these solutions.   At these values, the entire equation of motion \eqref{aeom} cancels identically except for the source term.  I.e., we are left with the inconsistent expression 
\be
0  =\frac{8\pi G\,\mu_0}{a^3} \, .
\ee
It was shown in \cite{deRham:2013wv} that cosmology in the candidate partially massless model is pure gauge: in the absence of a source term, any arbitrary function is a solution for the scale factor $a$.  When coupled to a source, however, we see there is no solution.

\vspace{.5cm}
\noindent {\bf Non-Minimal Model: $\beta_2 \neq 0$ and $\beta_3 \neq 0$.}   For totally generic $\beta_2$ and $\beta_3$, it becomes challenging to extract the physical roots of \eqref{aeom} and thus find $f[a]$ and the corresponding expressions for $V_1(R_0)$ and ${\cal R}(R_0^+)$.  However, the expressions simplify greatly if we consider the ${\mathbb Z}_2$ model with $\beta_0 = \beta_4$ and $\beta_1 = \beta_3$ and set $m^2 = 2 H^2$.  This choice leaves one parameter of the theory undetermined which we parametrize by $\beta_3$.  Consistent with the constraints found above, we assume $\beta_3 < 0$ and express our results in terms of the absolute value $|\beta_3|$.  We find 
\be
f[a] = -\frac{H}{k} a^2 (1-a)\left(2-\frac{H}{k} (1-a)\right) ,
\ee
and
\be
k=\left(\frac{3H GM}{|\beta_3| r_0^3}\right)^{1/3}+H \, .
\ee
We see immediately that this choice of parameters no longer recovers General Relativity when $m,H \rightarrow 0$.  Furthermore, we find that $f[a]$ becomes negative when $a<1$.  Consistency with our initial conditions \eqref{init} then forces us to choose the positive root:
\be
a'=+k\sqrt{f[a]} \, .
\ee
Thus, for these parameters we find an expanding solution, indicating that a black hole will not form.  Indeed, the expression for the putative apparent horizon is now given by
\be
V_1(R_0 \!= \!R_{hor}) =1 - \left(H R_{hor}+ \left(\frac{3 H G M}{|\beta_3|}\right)^{1/3} \right)^2=0 \, .
\ee
This expression only has one positive root which corresponds to the de Sitter horizon $\sim \tfrac{1}{H}$ and not to a black hole horizon.  %Generalizing to $m^2 \geq 2 H^2$ does not alter this conclusion.

\section{Vacuum Solution Matching}
\label{matching}
 \vspace{-.5cm}
Of the points in parameter space considered, only the minimal model contained a collapsing solution that we might reasonably expect gives rise to a black hole and has a smooth limit with General Relativity when $m,H \rightarrow 0$.   We now consider the matching of this solution to an exterior vacuum solution.  In particular, we want to know whether the metric for the collapsing ball of dust can be matched to a static solution on the exterior.  The picture would be analogous to that of General Relativity in which a collapsing star ``uncovers" the static Schwarzschild metric as it collapses.

A generic static, spherically symmetric metric can be written as
\be
\label{gout}
g_{\mu\nu} dx^\mu dx^\nu = -V_0(R)dT^2+ \frac{dR^2}{V_1(R)}+R^2 d\Omega^2\, ,
\ee
for some coordinates $T$ and $R$.  If we assume that the metric outside the star takes this form, then $V_0(R)$ and $V_1(R)$ are fixed entirely by matching at the surface of the star.  In particular, for the minimal model, expression \eqref{V1gen} requires
\begin{align}
\label{V1out}
V_1(R) = &1-\frac{2\,GM}{R} - H^2 R^2
-\frac{m^2R^2}{2H^2}\left(m^2-2H^2 \right) \nonumber \\
&-\frac{m^2R^2}{H} \sqrt{\frac{2GM}{R^3} +\frac{\left(m^2-2H^2 \right)^2}{4H^2}}  \, ,
\end{align}
outside of the star.  We now check whether this is a valid solution to the vacuum equations of motion that follow from \eqref{L}.

To do so, we start by allowing the reference metric $f_{\mu\nu}$ to be de Sitter spacetime in a completely arbitrary coordinate system.  As before, the coordinate system will be fixed by the equations of motion.  Using the equation \eqref{gout} for the $g$-metric, we find that the equations of motion enforce that $f_{\mu\nu}$ be diagonal with $g_{\mu\nu}$ and also independent of the time coordinate $T$.  The most generic ansatz for $f_{\mu\nu}$ that satisfies these constraints is
\begin{multline}
f_{\mu\nu} dx^\mu dx^\nu =\\
 -\left(1-H^2 P(R)^2\right) dT^2+ \frac{P'(R)^2\, dR^2}{1-H^2 P(R)^2}+P(R)^2 d\Omega^2\, ,
\end{multline}
for an arbitrary function $P(R)$.

The resulting equations of motion give three equations for the three functions $V_0(R)$, $V_1(R)$ and $P(R)$:
\begin{multline}
\frac{V_1'(R)}{R}+\frac{V_1(R)-1}{R^2}+3 H^2\\
+m^2\left(\frac{\sqrt{V_1(R)}P'(R)}{\sqrt{1-H^2P(R)^2}}+\frac{2 P(R)}{R}-3   \right) = 0 \, ,
\end{multline}
\begin{multline}
\frac{V_0'(R)}{RV_0(R)}V_1(R)+\frac{V_1(R)-1}{R^2}+3 H^2\\
+m^2\left(\frac{\sqrt{1-H^2P(R)^2}}{\sqrt{V_0(R)}}+\frac{2 P(R)}{R}-3   \right) = 0 \, ,
\end{multline}
\begin{multline}
\frac{V_0'(R)}{RV_0(R)}+\frac{4}{R^2}\left(1-\frac{\sqrt{1-H^2P(R)^2}}{\sqrt{V_1(R)}}\right)\\
+\frac{2H^2P(R)}{R\sqrt{V_0(R)V_1(R)}}= 0 \, ,
\end{multline}
where primes now denote derivatives with respect to $R$.  While finding an exact analytic solution to these equations is challenging, it is straightforward to show that \eqref{V1out} is not a valid solution for $V_1(R)$, for any $V_0(R)$ and $P(R)$.  Moreover, because the $g$- and $f$-metrics are static and simultaneously diagonal, the vacuum solution that solves the above equations will inevitably possess an infinite curvature singularity at the horizon \cite{Deffayet:2011rh}, in tension with \eqref{Rgen}.  The failure of the collapsing star solution to match with a static vacuum solution on the boundary provides evidence that the physical black hole solutions are likely time-dependent.

\section{Discussion}
\label{discuss}
\vspace{-.5cm}
By considering gravitational collapse for massive gravity on de Sitter spacetime we have seen how the Vainshtein mechanism works differently in de Sitter, with the minimal model recovering the solutions of General Relativity when $m\rightarrow 0$, in contrast to the minimal model in flat spacetime.  Moreover, we've seen that, for these solutions, a smooth flat spacetime limit $H \rightarrow 0$ also necessitates taking $m\rightarrow 0$.  This might have been expected: the adoption of flat asymptotics from the outset along with the assumption of a homogenous and isotropic collapse \eqref{gmetric} would lead to inconsistent equations of motion.  Thus, the solutions presented here are not useful for studying gravitational collapse or black hole solutions with flat asymptotics in massive gravity.  

Nevertheless, the de Sitter solutions are interesting in their own right.  The solutions found here for the minimal model describe a physically plausible collapse that results in a black hole with finite curvature at the apparent horizon and a necessarily time-dependent vacuum metric outside the star.   Because these solutions are arbitrarily close to those of General Relativity in the limit of small graviton mass,  it is not obvious from this analysis that there would be any obstruction to forming physical black holes within the regime of validity of the massive gravity effective theory.

The fact that not all points in parameter space give rise to collapsing solutions is perhaps not surprising.  If moved to the r.h.s. of equation \eqref{EOM}, the graviton mass term can be thought of as an exotic fluid whose properties, like a cosmological constant, might prevent collapse.  Our analysis thus indicates that there may be obstructions to finding physical black hole solutions for all possible parameters of the ghost-free massive gravity theory.

Finally, we note that the issue of superluminalities around spherically symmetric solutions in massive gravity remains an open question.  The solutions presented here might be used to better understand concretely whether any causality violation exists.

%The central caveat to our analysis is that it's still not obvious whether we are solving on the correct physical branch.  The assumptions of equation \eqref{tr} greatly simplify our expressions and allows us to solve otherwise intractable equations analytically.  However, it's possible that other physically reasonable solutions exist away from this assumption.  Nevertheless, the solutions found here for the minimal model describe a physically plausible collapse that results in a black hole with finite curvature at the apparent horizon and a necessarily time-dependent vacuum metric outside the star.

\vskip.5cm

\bigskip
\noindent {\bf Acknowledgements}: 
We would like to thank Claudia de Rham, Kurt Hinterbichler and Laura Johnson for discussions and comments on the draft. This work was supported by DOE grant DE-SC0011941 and Simons Foundation Award Number 555117.

\bibliographystyle{apsrev4-1}
\bibliography{collapse_arxiv}

\end{document}